\documentclass[12pt,letterpaper]{article}
\usepackage{amsmath,amssymb}
\usepackage{graphicx,color}
\usepackage[linkcolor=blue,colorlinks=true]{hyperref}
\usepackage{cite,epsf}
\usepackage[bf]{caption}
\usepackage{subfig}
\usepackage{mathrsfs}
\usepackage[nottoc,notlot,notlof]{tocbibind}

\def\a{\alpha}

\def\be{\begin{equation}}
\def\ee{\end{equation}}
\def\bea{\begin{eqnarray}}
\def\eea{\end{eqnarray}}
\def\ba{\begin{aligned}}
\def\ea{\end{aligned}}

\def\12{\frac{1}{2}}

\usepackage{xargs}
\usepackage[colorinlistoftodos,prependcaption,textsize=tiny]{todonotes}
\newcommandx{\Stefan}[1]{\todo[backgroundcolor=red!25,bordercolor=blue,noline]{St:#1}}
\newcommandx{\Sadik}[1]{\todo[backgroundcolor=blue!25,bordercolor=red,noline]{Sa:#1}}

\numberwithin{equation}{section}

\begin{document}

\begin{titlepage}       \vspace{10pt} \hfill 

\vspace{20mm}

\begin{center}

{\large \bf The Third Way to Interacting $p$-form Theories}

\vspace{30pt}

Matteo Broccoli$\,{}^{a}$, Nihat Sadik Deger$\,{}^{b}$ and Stefan Theisen$\,{}^{a}$
\\[6mm]

{\small
{\it ${}^a$Max-Planck-Insitut f\"ur Gravitationsphysik, Albert-Einstein-Institut\\
Am M\"uhlenberg 1, D-14476 Potsdam, Germany}\\[2mm]
{\it ${}^b$Department of Mathematics, Bogazici University,\\ 
Bebek, 34342, Istanbul, Turkey}
}

\vspace{20pt}

\end{center}

\centerline{{\bf{Abstract}}}
\vspace*{5mm}
\noindent
We construct a class of interacting $(d-2)$-form theories in $d$ dimensions that are 
`third way' consistent. This refers to the fact that the interaction terms in the 
$p$-form field equations of motion neither come from the variation of 
an action nor are they off-shell conserved on their
own. Nevertheless the full equation is still on-shell consistent.
Various generalizations, e.g. coupling them to $(d-3)$-forms, where   
3-algebras play a prominent role, are also discussed. 
The method to construct these models  also easily recovers    
the modified 3$d$ Yang-Mills theory obtained earlier 
and straightforwardly allows for higher derivative extensions.

\vspace{15pt}

\end{titlepage}

\section{Introduction}

Study of three-dimensional gravity theories brought out a new class of   
theories, labelled as ``third way consistent",
whose field equations contain tensors whose divergences   
vanish only on-shell \cite{Bergshoeff:2015zga}.
These models do not have actions which contain only the metric field and the 
Bianchi identity alone does not guarantee the consistency of the equations of motion. The first such example was
the Minimal Massive Gravity (MMG) \cite{Bergshoeff:2014pca} which   
avoids the clash between bulk and boundary unitarity which is a feature of generic 
three-dimensional higher derivative gravity theories; in its parameter space   
there is a region where both left and right central charges of the   
dual CFT and the energy of the massive bulk graviton mode are positive \cite{Bergshoeff:2018luo}. These desirable features motivated   
further investigation of this type of theories and
more gravity examples were constructed \cite{Ozkan:2018cxj,   
Afshar:2019npk}. But this structure is not unique
to gravity, in \cite{Arvanitakis:2015oga} a Yang-Mills theory with   
this property was found.\footnote{See also \cite{Mukhi:2011jp,Nilsson} where special cases of the 
systems discussed in \cite{Arvanitakis:2015oga} had been constructed 
previously.} 
Since all these are in
three dimensions, it is natural to ask whether such models exist in higher   
dimensions. In this paper we answer this question affirmatively for $p$-form theories.

Our construction is based on shifting a flat Yang-Mills gauge   
connection with a Lie algebra valued 1-form. The resulting equation   
from the vanishing field strength condition is third-way consistent   
thanks to the Jacobi identity and cannot be derived from an action.   
This shifting idea is inspired by the fact that in the first order   
formulation of the third way consistent gravity examples   
\cite{Bergshoeff:2014pca, Ozkan:2018cxj, Afshar:2019npk} their spin   
connection is similarly shifted. In the next section we first show   
that in $d=3$ choosing the
shift appropriately it is possible to recover the model constructed in   
\cite{Arvanitakis:2015oga}. With this approach finding extensions of   
\cite{Arvanitakis:2015oga} becomes straightforward and systematic for   
which we give one example. In section 3,
we study this issue in general $d$-dimensions. If the shifting 1-form   
is a dual field strength of a $(d-2)$-form, then one obtains a theory   
of $(d-2)$-forms coupled Yang-Mills gauge fields. However, requiring   
the compatibility of the gauge symmetries enforces to set the   
Yang-Mills field to zero and one ends up with an interacting   
$(d-2)$-form theory, which in $d=3$ corresponds to the ungauged   
version of
\cite{Arvanitakis:2015oga}. $p$-form theories have been studied   
extensively in the past and they are important ingredients of supergravity and string theory. References 
to some of the early work include 
\cite{Freedman:1977pa, Townsend:1979hd, Freedman:1980us,   
ThierryMieg:1982un, Blau:1989bq, Quevedo:1996uu,Brandt:1997ws,Henneaux:1997ha,   
Henneaux:1997bm}.
The novelty of our model is that interaction is possible without   
modifying the standard $p$-form gauge transformation unlike, for   
example, the well-known 4-dimensional Freedman-Townsend (FT) model with   
2-forms \cite{Freedman:1977pa, Townsend:1979hd, Freedman:1980us}. Although, the field equation
of this model has the same form as ours, namely a flatness condition, theirs comes from a non-polynomial action.
We also discuss extensions of our model with additional fields and in particular with
$(d-3)$-forms in which 3-algebras enter when one imposes the consistency condition. 
We conclude in Section 4 with some future directions.

\section{Shifting the connection}

In this section we review the third way consistent three dimensional gauge theories which were found in \cite{Arvanitakis:2015oga}. 
We present them in a way which can be easily generalized. Some generalizations will be presented in this section, others, 
in particular those to higher dimensions with higher form gauge field, will be the subject of the next section. 

The starting point is a flat connection of an arbitrary gauge group $G$, i.e. a connection which solves the 
equation 
\be
F(A)=dA+A^2=0 \, .
\label{cs}
\ee
As it simplifies the calculations, we will mostly use form notation. Here $A=A_\mu^i T_i dx^\mu$ 
is the gauge field one-form and $i=1,\dots,{\rm dim}(G)$ is a gauge index. The $T_i$ are (anti-hermitian) 
representation matrices which satisfy 
$[T_i,T_j]=f^k{}_{ij}T_k$. The structure constants satisfy the Jacobi-identity 
$f^i{}_{j\underline{k}}f^j{}_{\underline{mn}}=0$,
where we anti-symmetrize over the underlined indices.
We will be mostly considering compact semi-simple gauge groups where we can 
lower the first index with the Cartan-Killing metric such that $f_{ijk}$ is totally anti-symmetric. 

Consider now an arbitrary Lie-algebra valued one-form $C=C^i_\mu T_i dx^\mu$ and the equation 
\be\label{F(AB)}
F(A+C)=F(A)+D_A C+C^2=0 \, ,
\ee
where $D_A C=dC+AC+CA$ is the covariant derivative of $C$.

For consistency, eq.\eqref{F(AB)} should satisfy $D_A F(A+C)=0$. While $D_AF(A)\equiv0$ (Bianchi identity), we now find the 
condition 
\be\label{con1}
D_A^2 C+D_A C^2=[F(A),C]+[D_A C,C]=0 \, ,
\ee
where we have used $D_A^2 C=[F(A),C]$ and $D_A C^2=[D_A C,C]$. Using the equation of motion \eqref{F(AB)}
and $[C^2,C]=0$, we find that \eqref{con1} is indeed satisfied on-shell. 

In  three dimensions a simple generalization of \eqref{F(AB)} exists when $C$ is conserved.
In this case 
\be\label{F(AB)t}
F(A+C)+\tau *\!C=0 \qquad\hbox{with}\qquad D_A *\!C=0 \, ,
\ee
is also on-shell consistent, using $[*C,C]=0$. Here $\tau$ is an arbitrary constant parameter. 
Equations \eqref{F(AB)} and \eqref{F(AB)t} are gauge covariant if $C$ transforms under gauge transformations as 
$F$, i.e. $C \to g^{-1} C g$, $g\in G$.

Note that our starting point, namely \eqref{cs}, is the equation of motion following from the Chern-Simons action. 
For the choices of $C$ which we will discuss below, 
and which depend on the field strength $F(A)$,  this is not the case for \eqref{F(AB)} and \eqref{F(AB)t}.
In other words, they cannot be derived from a local action unless one introduces auxiliary fields similar 
to the model discussed in \cite{Arvanitakis:2015oga}. 
If they could be derived from an action, consistency 
would be automatic.  

We now consider some special choices for $C$. This recovers models constructed in \cite{Arvanitakis:2015oga}, but also 
new third way consistent theories. 

\begin{enumerate}

\item
The choice $C=\kappa *\!F(A) \equiv \kappa\, \tilde F$, which satisfies $D_A *C=0$; $\kappa$ is an arbitrary parameter
of dimension 1/mass.  
This from \eqref{F(AB)t} leads to the third way consistent equation 
\be
(1-\kappa\,\tau)F+\kappa\,D \tilde F+\kappa^2\,\tilde F^2=0 \, ,
\ee
where here and below $D=D_A$. In components this becomes
\be\label{Town1}
\varepsilon^{\mu\nu\rho}\left(D_\nu\tilde F_\rho+\frac{1}{2m}\tilde F_\nu\times\tilde F_\rho\right)+\mu\,\tilde F^\mu=0 \, ,
\ee 
To make contact with the notation used in \cite{Arvanitakis:2015oga}, we have defined 
$\tilde F_\mu\times\tilde F_\nu=[\tilde F_\mu,\tilde F_\nu]$, $\kappa=1/m$ and $\tau=m-\mu$, 
where $m$ and $\mu$ are mass parameters. For this choice of $C$, the shift $A \to A+\kappa *\!F(A)$ 
was used as a parity transformation of the gauge field $A$ in \cite{Arvanitakis:2015oga}. 
For the special case $m=2\mu$ this system had already appeared in \cite{Nilsson}. 

\item
If we use instead $C=\kappa\tilde F+\a\,j$, where $j$ is a conserved matter current, i.e. $D*j=0$ and, therefore, 
$D*\!C=0$, we find from \eqref{F(AB)t} the equation
\be
F(A+\kappa\tilde F+\a j)+\tau *\!(\kappa\tilde F+\a j)=0 \, .
\ee
To make contact with ref.\cite{Arvanitakis:2015oga} we define $\kappa=1/m,\,\tau=m-\mu$ and normalize the matter 
current by choosing $\a=\kappa/\tau$. We then find the following component expression 
\be
\varepsilon^{\mu\nu\rho} \left(D_\nu\tilde{F}_\rho + \frac{1}{2m} \tilde{F}_\nu \times \tilde{F}_\rho \right) \, 
+ \mu \tilde{F}^\mu = J^\mu \, ,
\label{Town2}
\ee
where 
\be
J^\mu= j^\mu - \frac{\varepsilon^{\mu\nu\rho}}{m-\mu}\left( D_\nu j_\rho + 
\frac{\tilde{F}_\nu \times j_\rho}{m} + \frac{j_\nu \times j_\rho}{2m(m-\mu)}\right) \, .
\label{Town3}
\ee
This agrees\footnote{Up to the sign of the last term.} with ref.\cite{Arvanitakis:2015oga}, but it has now been derived without 
any effort. Note that
any choice of $C$ in \eqref{F(AB)} which is not conserved would result in a third way
consistent model that is not of the form \eqref{Town2}.

\item
So far we have shown how to obtain the models constructed in ref.\cite{Arvanitakis:2015oga} by the method of 
`shifting the connection'. Since this method is very general, we can easily construct new third way consistent 
gauge models in three dimensions. For instance the choice $C=\kappa *D\tilde F$ in \eqref{F(AB)t}, which satisfies 
$D*C=\kappa D^2\tilde F=\kappa[F,\tilde F]=0$, leads to\footnote{In odd dimensions with Minkowski signature, $**=-1$ for any $p$-form.} 
\be
F(A+\kappa *D\tilde F)-\tau\kappa\,D\tilde F=0 \, .
\ee
In components we find
\be
\varepsilon^{\mu\nu\rho}\left(-\tau\, D_\nu\tilde{F}_\rho +  \frac{1}{2}\tilde{F}_\nu \times \tilde{F}_\rho
+\frac{1}{2}\kappa\,\varepsilon_{\nu\alpha\beta}\varepsilon_{\rho\sigma\delta} D^\alpha\tilde{F}^\beta\times
D^\sigma\tilde{F}^\delta \right)
+ \frac{1}{\kappa}\tilde{F}^\mu
=  - D^2\tilde F^\mu \, ,
\label{higher}
\ee
which is a higher derivative extension of \eqref{Town1}. We remark that the coefficient of $\tilde{F}^\mu$ cannot 
be set to zero unlike in \eqref{Town2}.

\end{enumerate}

\section{The third way with $p$-forms}

The discussion so far was in three spacetime dimensions. An immediate question is whether the construction given there, which 
uses a shifted connection, also works dimensions other than three. This is indeed the case.     
In $d\geq 3$ dimensions we choose $C$ as the 
dual of a Lie algebra valued $(d-1)$-form $H$, i.e. $C\equiv \kappa\,\tilde H=\kappa *\!H$ with $H=H^i T_i$; $\kappa$ is a 
coupling constant. From $F(A+\kappa\tilde H)=0$ we obtain the equation of motion
\be
F(A)^i+\kappa\, D_A \tilde{H}^i = -\frac{1}{2} \kappa^2 f^i{}_{jk}\,\tilde{H}^j\wedge \tilde{H}^k \, ,
\label{gauged}
\ee
which is the $d$-dimensional version of \eqref{F(AB)} that we considered 
above.\footnote{Note that for $d\neq 3$ there is no analogue of \eqref{F(AB)t}. Moreover, the choice $C=*F$ is not available either. 
Hence, in higher dimensions our method does not lead to a third way extended version of the standard Yang-Mills theory.  
But, choosing 
$C=*D*F$ is possible in all dimensions which will give rise to a modified flatness condition with higher derivative terms.
}
As is stands, the equation is Yang-Mills gauge covariant if $\tilde H^i$ transforms as $\delta_\lambda \tilde H^i=f^i{}_{jk}\tilde H^j \lambda^k$
under $\delta_\lambda A^i=d\lambda^i+f^i{}_{jk}A^j \lambda^k$. 

We now want to elevate $H^i$ to a dynamical field. To this end we define it to be the Yang-Mills covariant field strength of 
a $(d-2)$-form $B^i$, i.e. $H^i=(DB)^i=d B^i+f^i{}_{jk}A^j B^k$. $B^i$ transforms covariantly under Yang-Mills transformations, 
i.e. like $H^i$.  
Being the curl of a $p=(d-2)$-form, we would like $H^i$ to have the $p$-form 
gauge symmetry. To make it compatible with the Yang-Mills symmetry, we need to define the $p$-form transformation rules 
for $B^i$ and $A^i$ as 
\be
\delta_\xi B^i=D\xi^i=d\xi^i+f^i{}_{jk}A^j\xi^k\,,\qquad \delta_\xi A^i=0 \, ,
\ee
where $\xi^i$ are $(d-3)$-forms. But this implies
\be
\delta_\xi H^i=f^i_{jk}F^j\xi^k \, .
\ee
Covariance of $H^i$ and of the equations of motion
require $F(A)=0$.\footnote{Including the term $f^i{}_{jk}\tilde H^j\xi^k$ in the transformation of $B^i$
as in the FT-model \cite{Freedman:1980us} does not change this conclusion. Also note that $d=3$ case is exceptional where $B^i$
and $A^i$ can be identified.}
The simplest way to achieve this is by setting $A=0$. This leaves an equation which involves only 
$H^i=dB^i$ in \eqref{gauged}, namely
\be\label{eom1}
d\tilde H^i=-\tfrac{1}{2}{\kappa}  \, \tilde H^j\wedge\tilde H^k\,f^i{}_{jk} \, .
\ee
This is a system of second order equations for a collection of interacting $(d-2)$-forms $B^i$.  
If the coupling constant $\kappa=0$, we get a collection of free $(d-2)$-forms.

Third way consistency of the deformed system
follows from the construction and can be easily checked explicitly: acting with the exterior derivative, 
the left hand side vanishes identically due to $d^2=0$, but the right hand side vanishes only on-shell, after using the 
Jacobi identity. Obviously $H^i$ is invariant under $\delta B^i=d\xi^i$. The system of equations has a global symmetry under $G$, 
under which $B^i$ transforms as before, but with 
constant parameters $\lambda^i$. However, gauging this symmetry fails for the reasons given above. 

It is not difficult to show that the equations of motion cannot be obtained as the variation of an 
action which does not contain additional (auxiliary) fields. The equations are of the form
$\mathscr{L}(B)^i_{\underline{\mu}}=0$,  
and one needs to check the integrability condition, i.e. whether 
$\frac{\delta}{\delta B^j_{\underline{\nu}}(y)}\mathscr{L}(B)^i_{\underline{\mu}}(x)-(x\leftrightarrow y,i\leftrightarrow j,
\underline{\mu}\leftrightarrow\underline{\nu})$ vanishes. If it doesn't, then there is no action 
from which the equations of motion follow. In this way one verifies that neither \eqref{eom1} nor the other 
equations in this section can be derived from an action which does not contain additional, auxiliary fields.    

Note that \eqref{eom1} has the form of a flatness condition\footnote{
We may also construct higher derivative
extensions of this model by adding further terms in the shift $C$, e.g. $\kappa*d*(\tilde H \wedge \tilde H)$. The equation 
of motion \eqref{flatness} will be modified to $F(\tilde H + \kappa*d*(\tilde H \wedge \tilde H) )=0$. As long as the shift 
is built from the field strength $H$, $p$-form symmetry is guaranteed to work off-shell.}
\be
F(\tilde H)=0 \, ,
\label{flatness}
\ee
pretending that $\tilde H$ is a gauge field. From this it follows that $\tilde H$ is pure gauge, i.e. 
\be
\tilde H =U^{-1}d U
\ee
where $U$ is a group element and 
where we have set $\kappa=2$ for simplicity. The Bianchi identity is $d H=0$. 
This immediately reminds the principal chiral sigma model (for a review see e.g. \cite{Lu:2008kb}) which is defined by the Lagrangian 
\be
\label{sigma}
L= \textrm{Tr}(\partial_\mu \hat U^{-1} \partial^\mu\hat U) \, ,
\ee 
where $\hat U$ is an arbitrary Lie group element. One can define a connection 1-form that takes values in the corresponding Lie algebra 
as $\hat{A}= \hat U^{-1}d \hat U$ and then the equations of motion derived from \eqref{sigma} 
state that $\hat A$ is co-closed, i.e. $d*\hat{A}=0$. 
Moreover, flatness $F(\hat{A})= d\hat{A} + \hat{A} \wedge \hat{A}=0$ is satisfied identically, as $\hat A$ is pure gauge. 
We therefore see that that conditions which
$\hat{A}$ and $\tilde H$ satisfy are the same, but with the role of the equations of motion and the  Bianchi identity 
reversed. In this sense the two models are dual to each other. 

This brings us to the comparison of our third way consistent model of $p=(d-2)$ forms $B^i$ with the Lagrangian 
model of Freedman and Townsend \cite{Freedman:1980us} in $d=4$.
They also considered interacting Lie-algebra valued anti-symmetric tensor fields $B^i$, which are coupled to a two-form current 
$j=*(*dB^i\wedge *dB^i)$. Minimal coupling, which modifies the equations of motion, destroys current conservation and requires 
further non-minimal couplings, eventually leading to a non-polynomial action and equations of motion. 
Current conservation guarantees the invariance of the 
action under a deformed, field dependent $p$-form symmetry and the commutator of two variations on $B$
vanishes on-shell.\footnote{The composite field $*dB^i+{\cal O}(B^2)$ 
of Freedman and Townsend (called $A^i$ there) 
can be obtained from the  definition $H=D B$ if $D$ is a covariant derivative with `gauge field' 
$\tilde H^i$ and solving for $\tilde H$.  Their $\hat A$ is obtained if one uses a covariant derivative with connection $\tilde H+v$ 
instead, where $v$ is an independent gauge field (which we 
called $A$ in \eqref{gauged}).} 
The equations of motion in the FT-model reduce to 
the ones we have been considering, namely \eqref{flatness}, if one formally identifies their composite `gauge field' $A^i$ with our $\tilde H^i$. This amounts to dropping all higher order terms in $A^i$, which has an infinite power series expansion in the $p$-form field $B^i$. 
The model is dual to the principal chiral 
model in the sense that both can be derived from the same first order action. 
However, the equation of motion of the principal chiral model, $\partial^\mu A_\mu^i=0$, is not satisfied identically 
in the dual FT model, but it holds on-shell.

\medskip

We now discuss various generalizations of our third way consistent system of equations \eqref{eom1}, which consist of 
including additional fields.  

\medskip

\noindent
{\it Coupling to gravity}

To couple the third way consistent system of $(d-2)$-forms to an external gravitational field is trivial. In that case 
\eqref{eom1} becomes
\be\label{eom_cov_1}
\nabla_\mu\tilde H^i_\nu-\nabla_\nu\tilde H^i_\mu=-\,\kappa\,f^i{}_{jk}\tilde H^j_\mu\,\tilde H^k_\nu \, .
\ee 
Due to the symmetry of the Christoffel symbols, they drop out and 
the only appearance of the metric is through the Hodge star implicit in $\tilde H^i$. 
On-shell consistency can easily be shown for \eqref{eom_cov_1}. We act with $\nabla_\rho$ and anti-symmetrize in all three indices. 
The l.h.s is identically zero:  we replace the product of the two covariant 
derivatives by half of their commutator and use the Bianchi identity
$R_{\underline{\mu\nu\rho}}{}^\sigma=0$ to show that it vanishes. The on-shell vanishing of the r.h.s. is as in the flat case.

If gravity is dynamical, we need to construct 
an energy-momentum tensor which is on-shell conserved. A possible candidate is $T_{\mu\nu}=\tilde H^i_\mu\tilde H^i_\nu-\frac{1}{2}
g_{\mu\nu}\tilde H^{i\, \rho}\tilde H^i_\rho$ \cite{Arvanitakis:2015oga}. 
On-shell conservation is straightforward to demonstrate. 

\medskip

\noindent
{\it Adding lower form fields}

A simple generalization, which is inspired by \cite{Brandt:1997ws,Henneaux:1997ha}, is to add $p$-forms $b^a$ 
(with $p<d-2$) whose field strengths and 
dual field strengths we denote by $h^a$ and $\tilde h^a$ respectively. They come with their own gauge symmetry. The following system of 
equations is easily shown to be third way consistent
\be
d\tilde H^i=-\tfrac{1}{2}{\kappa} f^i{}_{jk}\tilde H^j\,\tilde H^k\,,\qquad 
d\tilde h^a=-\kappa\, t_{i~b}^a\tilde H^i\,\tilde h^b\,,
\ee 
provided $[t_j,t_k]=f^i{}_{jk}t_i$, i.e. $b^a$ is a vector in some representation of $G$. 

Other generalizations containing $p$-forms of degree lower than $d-2$ are also possible.
Consider, for instance, in addition to the $B^i$,  
a single $(d-3)$-form $b$ with field strength $h=db$ and dual field strength 
$\tilde h$. 
Postulate its equation of motion as 
\be\label{eom2}
d\tilde h=\tilde H^i\wedge\tilde H^j\wedge\tilde H^k\,f_{ijk} \, ,
\ee
together with \eqref{eom1}. 
Consistency now requires $d\big(\tilde H^i\wedge\tilde H^j\wedge\tilde H^k\,f_{ijk}\big)=0$.
Indeed, on-shell, we have 
\bea
d\big(\tilde H^i\wedge\tilde H^j\wedge\tilde H^k\,f_{ijk}\big)
=-\frac{3\kappa}{2}\,\tilde H^m\wedge\tilde H^n\wedge\tilde H^j\wedge\tilde H^k\,f^i{}_{mn}f_{ijk}=0 \, ,
\eea
with the help of the Jacobi identity, where we used \eqref{eom1}. 

A slightly less trivial generalization is to include a collection of $(d-3)$-forms $b^i$ and to impose the equations of motion
\be
d \tilde h^i=f^i{}_{jkl}\tilde H^j\wedge\tilde H^k\wedge\tilde H^l \, ,
\label{general}
\ee
where $f^i{}_{jkl}$ is totally antisymmetric in its lower indices. 
Consistency now requires
\be
0=d(f^i{}_{jkl}\tilde H^j\wedge\tilde H^k\wedge\tilde H^l)
=-\frac{3\kappa}{2}\,f^i{}_{jkl}\,f^j{}_{mn}\tilde H^m\wedge \tilde H^n\wedge\tilde H^k\wedge\tilde H^l \, ,
\ee
which is satisfied if 
\be\label{cond2}
f^i{}_{j\underline{kl}}\,f^j{}_{\underline{mn}}=0 \, .
\ee
A simple special solution of this condition is
\be
f^i{}_{jkl}=v^i\,f_{jkl} \, ,
\ee
where, as before, $f_{jkl}$ are the structure constants of a simple Lie algebra.   
Then \eqref{cond2} is satisfied by virtue of the Jacobi identity. 
$v^i$ is an arbitrary vector which transforms in the adjoint representation of the Lie algebra. 
If we choose it to have only one non-zero component, say $v^i=\delta^{i1}$, the system reduces to  
\eqref{eom2} plus a collection of free $(d-3)$-forms. 

A more interesting solution to \eqref{cond2} can be obtained as follows. 
The appearance of $f^i{}_{jkl}$ suggests a relation to three-algebras, whose essential features we briefly recall;  for review and references   
see \cite{deAzcarraga:2010mr} and \cite{Bagger:2012jb}. 
Denoting its generators by $T_i$, one has 
\be
[i,j,k]\equiv [T_i,T_j,T_k]=f^l{}_{ijk}T_l \, ,
\ee 
and the $f^i{}_{\underline{jkl}}$ satisfy the fundamental identity
\be
[i, j, [k, l, m]] = [[i, j, k], l, m] + [k, [i, j, l], m] + [k, l, [i, j, m]] \,.
\ee
Equivalently
\be\label{fund}
f^n{}_{klm}f^p{}_{ijn}=f^n{}_{ijk}f^p{}_{nlm}+f^n{}_{ijl}f^p{}_{knm}+f^n{}_{ijm}f^p{}_{kln} \, .
\ee
Following \cite{Gomis:2008uv}, one introduces a bi-invariant metric $h_{ij}$ on the three-algebra 
to lower the first index and arrives at $f_{ijkl}$, which is anti-symmetric in all its indices. 
Still following \cite{Gomis:2008uv} we choose it to be $h_{ij}={\rm diag}(-1,1,\dots,1)$. 
One then splits the indices into $i=(0,a,\phi)$ and chooses
\be
f_{0abc}=f_{\phi abc}=f_{abc}\,,\qquad f_{0\phi ab}=f_{abcd}=0 \, ,
\ee
and $f_{abc}$ are the structure constants of a compact Lie algebra. This solves the fundamental identity \eqref{fund}. 
Note that the above choices imply that the Cartan-Killing metric on the Lie-algebra is $\delta_{ab}$. 

We choose furthermore that all components of $f^i{}_{jk}$ vanish 
except for $f^a{}_{bc}$, which are totally anti-symmetric structure constants
$f^a{}_{bc}=f_{abc}$. With these choices one verifies that \eqref{cond2} is also satisfied. 

The vanishing of some of the components of the structure constants 
$f^i{}_{jk}$ and $f^i{}_{jkl}$ implies that $b^0+b^\phi$ is a free and completely decoupled field, 
while $B^0$ and $B^\phi$ satisfy source-free equations, but they appear as sources in the equation for $b^a$.

Instead of adding $(d-3)$-form fields we can add a set of $(d-q)$-forms $c^i$ with dual field strengths $\tilde k^i$
and postulate the equations of motion 
\be
d\tilde k^i=\tilde H^{j_1}\wedge\cdots\wedge\tilde H^{j_q}\,f^i{}_{j_1\dots j_q} .
\ee
Consistency now requires the following relation for the `structure constants': 
\be
f^i{}_{j\underline{j_2 \dots j_q}}f^j{}_{\underline{mn}}=0 \, .
\ee
We have not explored this possibility, i.e. the existence of such algebraic structures.

\section{Outlook}

In this paper, we constructed third way consistent interacting   
$(d-2)$-form theories in $d$-dimensions and studied some of
their properties. Third way consistency allows to have interactions   
without modifying the standard $p$-form gauge symmetry
but the price is the lack of an action. However, it might be possible   
to find an action by introducing auxiliary fields as
it happened in \cite{Arvanitakis:2015oga}. It would be useful to   
determine such an action which would allow studying its
dynamical couplings with other fields such as gravity   
\cite{Sezgin:1980tp, Freund:1981qw} which may lead to some interesting applications   
of these models. 
Another reason why an action is desirable is that it might be a viable starting 
point for the (path-integral) quantization of these models, a question which is otherwise not clear 
how to address.
Constructing supersymmetric versions of our model
is another open problem.

Higher order form fields are sources for branes. Adding $(d-3)$-brane sources to the equations of motion 
for $(d-2)$-forms and 
studying the resulting systems of branes would also be of interest.  

Constructing third-way interacting $p$-form theories where all 
transformations are compatible with non-trivial charges is another
direction worth exploring. To achieve this, the so-called embedding tensor formalism \cite{deWit1,deWit2}, 
which is used in the classification of gauged
supergravities and where several $p$-forms are required, might be useful. In $d=4$ 
only one- and two-forms are needed \cite{SezginWulff} and
hence this could be a suitable set-up to couple our model
covariantly to Yang-Mills fields.\footnote{We thank the referee for this suggestion.}

The 3-dimensional Yang-Mills model of \cite{Arvanitakis:2015oga} is   
closely related to non-Abelian Chern-Simons
theories considered in \cite{Mukhi:2011jp} where a novel Higgs   
mechanism was found. It would be interesting to
see if something similar happens for our model and try to explain the interaction terms we have 
as spontaneous breaking of a local symmetry as it was shown for the MMG model\cite{Bergshoeff:2014pca} 
in \cite{Chernyavsky:2020fqs}.

Finding third way consistent gravity theories in $d>3$ is another  
challenge. In $d=3$, using the first order formulation of
gravity was useful \cite{Bergshoeff:2014pca, Ozkan:2018cxj,   
Afshar:2019npk} but in higher dimensions the metric description might   
be more practical. Note that the Christoffel symbol $\Gamma^\nu_{\mu\rho}$ can also be   
thought of as the
$(\nu, \rho)$ component of a Yang-Mills field: $\Gamma^\nu_{\mu\rho} =   
(A_\mu)^\nu_{\,\,\, \rho}$ \cite{Jackiw:2004mn}
and hence to obtain such models we expect a shift in the Christoffel   
connection starting from the vacuum Einstein equation or a   
modification of it that comes from an action. Indeed, we checked that   
the MMG model \cite{Bergshoeff:2014pca} can be obtained via the shift:   
$ \Gamma^\nu_{\mu\rho} \rightarrow \Gamma^\nu_{\mu\rho} 
+ \kappa\,\varepsilon_\mu^{\,\,\, \nu\alpha} S_{\alpha\rho}$ where   
$S_{\alpha\rho}= R_{\alpha\rho}-\frac{1}{4}Rg_{\alpha\rho}$ is
the Schouten tensor. However, there is an important difference with   
the gauge theory case that we studied in this paper, namely not every
shift leads to a third way consistent theory. For example, doing this   
shift with the Ricci instead of the Schouten tensor does not work. 
We  hope to come back to these issues in the near future.

\section*{Acknowledgements}
NSD is grateful to the Albert-Einstein-Institute, Potsdam, where this work was initiated, for generous   
hospitality and financial support. 
MB is supported by the International Max Planck Research School for Mathematical and 
Physical Aspects of Gravitation, Cosmology and Quantum Field Theory.
We thank P. Townsend and E. Bergshoeff for feedback on the manuscript. 



\begin{thebibliography}{99}

\bibitem{Bergshoeff:2015zga}
E.~Bergshoeff, W.~Merbis, A.~J.~Routh and P.~K.~Townsend,
``The Third Way to 3D Gravity,''
Int. J. Mod. Phys. D \textbf{24} (2015) no.12, 1544015
[arXiv:1506.05949 [gr-qc]].

\bibitem{Bergshoeff:2014pca}
E.~Bergshoeff, O.~Hohm, W.~Merbis, A.~J.~Routh and P.~K.~Townsend,
``Minimal Massive 3D Gravity,''
Class. Quant. Grav. \textbf{31} (2014), 145008
[arXiv:1404.2867 [hep-th]].

\bibitem{Bergshoeff:2018luo}
E.~A.~Bergshoeff, W.~Merbis and P.~K.~Townsend,
``On-shell versus Off-shell Equivalence in 3D Gravity,''
Class. Quant. Grav. \textbf{36} (2019) no.9, 095013
[arXiv:1812.09205 [hep-th]].

\bibitem{Ozkan:2018cxj}
M.~\"Ozkan, Y.~Pang and P.~K.~Townsend,
``Exotic Massive 3D Gravity,''
JHEP \textbf{08} (2018), 035
[arXiv:1806.04179 [hep-th]].


\bibitem{Afshar:2019npk}
H.~R.~Afshar and N.~S.~Deger,
``Exotic massive 3D gravities from truncation,''
JHEP \textbf{11} (2019), 145
[arXiv:1909.06305 [hep-th]].

\bibitem{Arvanitakis:2015oga}
A.~S.~Arvanitakis, A.~Sevrin and P.~K.~Townsend,
``Yang-Mills as massive Chern-Simons theory: a third way to three-dimensional gauge theories,''
Phys. Rev. Lett. \textbf{114} (2015) no.18, 181603
[arXiv:1501.07548 [hep-th]].

\bibitem{Mukhi:2011jp}
S.~Mukhi,
``Unravelling the novel Higgs mechanism in (2+1)d Chern-Simons theories,''
JHEP \textbf{12} (2011), 083
[arXiv:1110.3048 [hep-th]].

\bibitem{Nilsson}
B.~E.~W.~Nilsson,
``Critical solutions of topologically gauged N = 8 CFTs in three dimensions,''
JHEP \textbf{04} (2014), 107
[arXiv:1304.2270 [hep-th]].


\bibitem{Freedman:1977pa}
D.~Z.~Freedman,
``Gauge Theories of Antisymmetric Tensor Fields,''
CALT-68-624.

\bibitem{Townsend:1979hd}
P.~K.~Townsend,
``Covariant Quantization of Antisymmetric Tensor Gauge Fields,''
Phys. Lett. B \textbf{88} (1979), 97-101.


\bibitem{Freedman:1980us}
D.~Z.~Freedman and P.~K.~Townsend,
``Antisymmetric Tensor Gauge Theories and Nonlinear Sigma Models,''
Nucl. Phys. B \textbf{177} (1981), 282-296.

\bibitem{ThierryMieg:1982un}
J.~Thierry-Mieg and L.~Baulieu,
``Covariant Quantization of Nonabelian Antisymmetric Tensor Gauge Theories,''
Nucl. Phys. B \textbf{228} (1983), 259-284.

\bibitem{Blau:1989bq}
M.~Blau and G.~Thompson,
``Topological Gauge Theories of Antisymmetric Tensor Fields,''
Annals Phys. \textbf{205} (1991), 130-172.

\bibitem{Quevedo:1996uu}
F.~Quevedo and C.~A.~Trugenberger,
``Phases of antisymmetric tensor field theories,''
Nucl. Phys. B \textbf{501} (1997), 143-172
[arXiv:hep-th/9604196 [hep-th]].

\bibitem{Henneaux:1997ha}
M.~Henneaux and B.~Knaepen,
``All consistent interactions for exterior form gauge fields,''
Physical  Review D \textbf{56} (1997), R6076
[arXiv:hep-th/9706119 [hep-th]].

\bibitem{Brandt:1997ws}
F.~Brandt and N.~Dragon,
``Nonpolynomial gauge invariant interactions of 1 form and 2 form gauge potentials,''
[arXiv:hep-th/9709021 [hep-th]].

\bibitem{Henneaux:1997bm}
M.~Henneaux,
``Consistent interactions between gauge fields: The Cohomological approach,''
Contemp. Math. \textbf{219} (1998), 93-110
[arXiv:hep-th/9712226 [hep-th]].

\bibitem{Lu:2008kb}
H.~Lu, M.~J.~Perry, C.~N.~Pope and E.~Sezgin,
``Kac-Moody and Virasoro Symmetries of Principal Chiral Sigma Models,''
Nucl. Phys. B \textbf{826} (2010), 71-86
[arXiv:0812.2218 [hep-th]].

\bibitem{deAzcarraga:2010mr}
J.~A.~de Azcarraga and J.~M.~Izquierdo,
``n-ary algebras: A Review with applications,''
J. Phys. A \textbf{43} (2010), 293001
[arXiv:1005.1028 [math-ph]].

\bibitem{Bagger:2012jb}
J.~Bagger, N.~Lambert, S.~Mukhi and C.~Papageorgakis,
``Multiple Membranes in M-theory,''
Phys. Rept. \textbf{527} (2013), 1-100
[arXiv:1203.3546 [hep-th]].

\bibitem{Gomis:2008uv}
J.~Gomis, G.~Milanesi and J.~G.~Russo,
``Bagger-Lambert Theory for General Lie Algebras,''
JHEP \textbf{06} (2008), 075
[arXiv:0805.1012 [hep-th]].

\bibitem{Sezgin:1980tp}
E.~Sezgin and P.~van Nieuwenhuizen,
``Renormalizability Properties of Antisymmetric Tensor Fields Coupled to Gravity,''
Phys. Rev. D \textbf{22} (1980), 301.

\bibitem{Freund:1981qw}
P.~G.~O.~Freund and R.~I.~Nepomechie,
``Unified Geometry of Antisymmetric Tensor Gauge Fields and Gravity,''
Nucl. Phys. B \textbf{199} (1982), 482-494.

\bibitem{deWit1}
B.~de Wit, H.~Nicolai and H.~Samtleben,
``Gauged Supergravities, Tensor Hierarchies, and M-Theory,''
JHEP \textbf{02} (2008), 044
[arXiv:0801.1294 [hep-th]].

\bibitem{deWit2}
B.~de Wit and H.~Samtleben,
``The End of the p-form hierarchy,''
JHEP \textbf{08} (2008), 015
[arXiv:0805.4767 [hep-th]].

\bibitem{SezginWulff}
E.~Sezgin and L.~Wulff,
``Supersymmetric Proca-Yang-Mills System,''
JHEP \textbf{03} (2013), 023
[arXiv:1212.3025 [hep-th]].

\bibitem{Chernyavsky:2020fqs}
D.~Chernyavsky, N.~S.~Deger and D.~Sorokin,
``Spontaneously broken $3d$ Hietarinta/Maxwell Chern\textendash{}Simons theory and minimal massive gravity,''
Eur. Phys. J. C \textbf{80} (2020) no.6, 556
[arXiv:2002.07592 [hep-th]].

\bibitem{Jackiw:2004mn}
R.~Jackiw,
``Fifty Years of Yang\textendash{}mills Theory and our Moments of Triumph,''
[arXiv:physics/0403109 [physics]].


\end{thebibliography}
\end{document}